\documentclass[twocolumn,final]{svjour3}
\usepackage[utf8]{inputenc}
\usepackage{graphicx}
\usepackage{subcaption}
\usepackage{amsmath}
\usepackage{color}
\usepackage{microtype}
\usepackage{bm}		   
\usepackage{amsfonts}  
\usepackage{mathtools} 
\usepackage{cuted}
\usepackage{tikz}	   
\usetikzlibrary{shapes.geometric, arrows, positioning, tikzmark}
\usepackage{nomencl} 
\usepackage{xcolor}
\usepackage{mathptmx}
\usepackage{dblfloatfix} 
\usepackage{algpseudocode}
\usepackage[ruled]{algorithm}
\usepackage[round]{natbib}
\usepackage[breaklinks=true]{hyperref}
\usepackage{breakurl}

\usepackage{etoolbox} 
\newcommand*{\doi}[1]{\href{http://dx.doi.org/#1}{doi: #1}}

\DeclareMathOperator*{\argmin}{arg\,min}

\title{Modular-topology optimization with Wang tilings: An application to truss structures}
\author{Marek Tyburec \and Jan Zeman \and Martin Doškář \and Martin Kru\v{z}\'{i}k \and Matěj Lepš}

\institute{Marek Tyburec\textsuperscript{1}\\
		  \email{marek.tyburec@fsv.cvut.cz}\\
		  ORCID: 0000-0003-0798-0948\\
		  \\
		  Jan Zeman\textsuperscript{1}\\
		  \email{Jan.Zeman@cvut.cz}\\
		  ORCID: 0000-0003-2503-8120\\
		  \\
		  Martin Do\v{s}k\'{a}\v{r}\textsuperscript{1}\\
		  \email{martin.doskar@fsv.cvut.cz}\\
		  ORCID: 0000-0001-7581-0567\\
		  \\
		  Martin Kru\v{z}\'{i}k\textsuperscript{2}\\
		  \email{kruzik@utia.cas.cz}\\
		  ORCID: 0000-0003-1558-5809\\
		  \\
		  Mat\v{e}j Lep\v{s}\textsuperscript{1}\\
		  \email{matej.leps@fsv.cvut.cz}\\
		  \\
		\textsuperscript{1}Czech Technical University in Prague, Faculty of Civil Engineering, Department of Mechanics, Th\'{a}kurova 7, 16629 Prague 6, Czech Republic\\
		\\
		\textsuperscript{2}Czech Technical University in Prague, Faculty of Civil Engineering, Department of Physics, Th\'{a}kurova 7, 16629 Prague 6, Czech Republic
}

\begin{document}

\maketitle
\begin{abstract}
Modularity is appealing for solving many problems in optimization. It brings the benefits of manufacturability and reconfigurability to structural optimization, and enables a trade-off between the computational performance of a Periodic Unit Cell (PUC) and the efficacy of non-uniform designs in multi-scale material optimization. Here, we introduce a novel strategy for concurrent minimum-compliance design of truss modules topologies and their macroscopic assembly encoded using Wang tiling, a formalism providing independent control over the number of modules and their interfaces. We tackle the emerging bilevel optimization problem with a combination of meta-heuristics and mathematical programming. At the upper level, we employ a genetic algorithm to optimize module assemblies. For each assembly, we obtain optimal module topologies as a~solution to a convex second-order conic program that exploits the underlying modularity, incorporating stress constraints, multiple load cases, and reuse of module(s) for various structures. Merits of the proposed strategy are illustrated with three representative examples, clearly demonstrating that the best designs obtained by our method exhibited decreased compliance: from $56\%$ to $69\%$ compared to the PUC designs.
\end{abstract}



\newpage

\section{Introduction}\label{sec:intro}

Modular structures, composed of repeated building blocks (modules), offer multiple appealing advantages over non-modular designs. These include more economical mass fabrication, increased production productivity~\citep{Tugilimana:2016:SOT}, and better quality control~\citep{Mikkola:2003:MMP}. In addition, modules facilitate structural reconfigurability, conversion among designs with considerably different structural responses~\citep{Nezerka2018}. Finally, the design of modular structures enables structural efficiency to be balanced with design complexity~\citep{Tugilimana2019}, which often arises in optimal structures~\citep{Kohn:1986:ODR}.

Our approach to designing modular structures follows recent successful applications of Wang tiles in compression and reconstruction of heterogeneous microstructures~\citep{Novak2012,Doskar2014,Antolin2019,Doskar2020}, where modularity suppresses artificial periodicity artifacts inherent to the periodic-unit-cell approach, e.g., \citep{Zeman:2007:FRM}. Here, we focus on the reverse direction: designing modular structures or materials composed of a~compressed set of modular \textsc{Lego}\textsuperscript{\textregistered}-like building blocks. In this endeavor, the concept of Wang tiles provides us with a~convenient mechanism for describing and controlling module types as well as their interface types. Our approach is explained in the simplest setting: the topology optimization of truss structures.

Below, we review recent developments in the modular design of truss structures (Section \ref{sec:SOTA_trusses}), and modular micro\-structures,  (Section~\ref{sec:SOTA_micro}). Finally, we discuss the benefits of our approach in Section \ref{sec:SOTA_aim}.

\subsection{Design of modular trusses}\label{sec:SOTA_trusses}

The optimal design of modular trusses appears to be a~new, to a~large extent unexplored, branch of structural optimization. In one of the pioneering works in this area of research, \citet{Tugilimana:2016:SOT} developed a~method to optimize the topology of a~single module as well as the module's spatially-varying rotations within the design domain. With this method, the optimization part relies on a~plastic design formulation~\citep{Dorn:1964:ADO} and thus provides lower-bound designs only as the kinematic compatibility is neglected. The need for acquiring elastic design formulation resulted in a~follow-up work \citep{Tugilimana:2017:CDM}, which proposed a~non-convex formulation allowing for multiple load cases, stress constraints, multiple module types, as well as the module reusability among structures. However, this formulation still requires a~manual definition of the module spatial distribution within the design domains. Therefore, \citet{Tugilimana2019} introduced a~two-level approach to optimize the topology of multiple modules as well as their spatial placement within a~structural design domain. While the lower-level formulation they proposed additionally extends itself to self-weight and local and global stability constraints, the upper-level simulated annealing with an adaptive neighborhood ensures dynamic grouping of modules.

Another approach, that of \citet{Zawidzki2019}, proposes a~bilevel optimization method to optimize the Truss-Z system \citep{Zawidzki2012}, a~modular pedestrian network. In the upper-level of this approach, the NSGA-II algorithm optimizes the module outer shape to provide geometrically versatile structures that can construct connected paths between pairs of access points. Lower-level optimization then employs simulated annealing with an adaptive neighborhood and minimizes the module weight in the sizing optimization of circular thin-walled sections with constraints on the von-Mises stresses and the Euler buckling ratio.

Limiting the number of unique cross-section areas of truss structures to improve constructability resulted in the use of dynamic grouping in topology optimization. In contrast to the former methods, where modules comprise multiple truss elements, this approach groups individual cross-sections into sets whose number and cardinalities are a~result from the optimization. For example, \citet{Shea1997} adopted the shape annealing approach with a~grouping criterion based on the optimized non-grouped cross-section areas, and \citet{Togan2008} employed genetic algorithms with grouping criteria based on the internal forces and on the slenderness ratios. In \citep{Lemonge2011}, a~genetic algorithm extended to multiple cardinality constraints was developed and used to solve the frame structure sizing optimization problem.

\subsection{Design of modular microstructures}\label{sec:SOTA_micro}

Distinguished modular and structural scales in the design of modular trusses evoke the standard multi-scale topology optimization for the design of (meta-)material microstructures. In such settings, theoretically optimal designs occur when microstructures vary pointwise in macro-scale design domains \citep{Rodrigues:2002:HOM}, i.e., with each macro-point associated with an independent module type.

Since scale separation hinders the use of single-scale methods because enormous resolution is required, the scales are usually (de-)coupled with the inverse homogenization approach \citep{Sigmund1994}. Not only does this setting introduce manufacturability issues due to discontinuous material distribution over microstructural cell interfaces (overcome recently, e.g., in \citep{Garner2019}), it also remains computationally expensive. On the other hand, the optimized structures tend to approach the theoretical limits of structural efficiency~\citep{Groen:2017:HTO}. Aiming to alleviate the computational burden, the periodic unit cell (PUC) approach---repeating a~single, possibly grad\-ed, microstructural cell throughout the entire design domain---is often adopted~\citep{Sigmund:1995:TMP,Liu:2008:OSH,Stromberg:2010:ALT,Liu:2017:AMD}; however, this significantly compromises the quality of optimized designs.

Both aforementioned methods represent two extreme cases: the multi-scale approach generates efficient structures but requires high computational resources; the PUCs are structurally inefficient yet computationally cheap. An intermediate method thus seems preferable for balancing computational demands with structural efficiency. Such an intermediate method emerges from modular design, as the number of unit cells is finite, and their number directly controls computational costs.

In this context, \citet{Sivapuram:2016:SMS} extended the multi-scale optimization approach by limiting the number of microstructures, but their approach required to predefine the spatial placement of cells and still lacked material continuity across their boundaries. The most general methods, published by \citet{Li2018} and \citet{Zhang2018a}, introduced concurrent approaches to simultaneously optimize spatial placement and the topology of a~finite set of modules while ensuring the material continuity with artificially predefined kinematical connectors.

\subsection{Aims and novelty}\label{sec:SOTA_aim}

As seen from the state-of-the-art review, the optimum design of modular structures is a~rapidly-evolving line of research for the structural optimization community. In this contribution, we consider this inherently two-level design problem in its original form; similarly to earlier studies \citep{Tugilimana:2016:SOT,Tugilimana:2017:CDM,Tugilimana2019}, we design optimized topologies of modules (lower level) and the assembly plan (upper level) concurrently. In contrast to the earlier contributions, however, we develop a~convex lower-level formulation for the topology optimization of truss modules, while still allowing for stress constraints, module reusability, and multiple load cases. 

Thanks to convexity, we can reach true global optima for coarse discretizations and assure that the algorithm avoids poor local optima for module topologies. We find this convex subproblem very important in this incipient phase of research because it provides a~rigorous answer regarding what is achievable. Unfortunately, this convexity does not translate into continuum topology optimization, which explains our choice for discretization with truss elements.

Second, we adopt the formalism of corner Wang tilings to describe the assembly plan of modules. This generalization \citep{Novak2012,Doskar2014,Doskar2016,Doskar2018,Doskar2020} of the PUC allows us to design compressed yet non-periodic assemblies, a~novel class of connectable and reusable \mbox{(micro-)}structures and materials. Contrary to \citep{Li2018,Zhang2018a}, our approach generates mechanically compatible structures fully automatically and avoids prescribing fixed cell interfaces.

The rest of the paper is structured as follows. In Section~\ref{sec:backgound}, we introduce the formalism of corner Wang tilings and recall a~convex elastic-design formulation for topology optimization of trusses. Subsequently, we extend this formulation with modularity, stress constraints, multiple load cases, and module reusability. To handle the discreteness of the modular assembly plan, a~genetic algorithm is applied as the solver in the upper-level optimization in Section~\ref{sec:methodology}. Finally, Section~\ref{sec:example} illustrates our method with three examples, assessing the scalability of the approach and multiple constraint types, and leads us to the conclusion that the proposed methodology is fairly efficient.

\section{Background}\label{sec:backgound}

\subsection{Wang tilings}\label{sec:vertex_based}

One of the goals of this article is to explore the merits of Wang tilings~\citep{Wang:1961:PTP} for the optimal design of modular structures. Wang tiles---unit squares with colored edges and fixed orientation---constitute a~formalism introduced by \citet{Wang:1961:PTP} to visualize the $\forall\exists\forall$ decidability problem of predicate calculus using an equivalent domino problem. Wang considered an infinite number of copies of an arbitrary set of Wang tiles and investigated whether there exists a~simply-connected tiling of the infinite plane such that the adjoining edges of neighboring tiles share the same color---the so-called valid tilings. Contrary to Wang's conjecture, \citet{Berger1966} built a~tileset that covers the infinite plane and that yet does not allow any periodic pattern to emerge, a~property proved by a~reduction from the Turing machine halting problem \citep{Turing1936}.

The two properties that made Wang tiles appealing in multi-disciplinary research are notably Turing completeness and the ability to form aperiodic patterns. While computations by self-assembly of DNA \citep{Winfree1998,Winfree2000} and automated theorem proving \citep{Wang:1961:PTP} exploit the Turing completeness, non-periodic patterns apply in computer graphics to construct compressed yet naturally-looking textures \citep{Cohen:2003:WTI}. The latter research also provided the motivation to employ Wang tiles in compression and reconstruction of microstructures, generalizing the concept of the periodic unit cell~\citep{Novak2012,Doskar2014,Doskar2016,Doskar2018,Doskar2020}.

Although traditional Wang tiles maintain information continuity across the edges, some discontinuity artifacts might appear in their corners. To solve this so-called corner problem~\citep{Cohen:2003:WTI} and avoid these periodically repeating artifacts~\citep{Doskar2020}, tiles with connectivity information stored in colored corners were proposed~\citep{Lagae:2006:AWT}. These corner tiles form a~subset of Wang tiles, as each combination of vertex color codes denotes a~unique edge type. Note that a~reverse procedure is not generally applicable \citep{Doskar2020}.

\begin{figure}[t]
	\includegraphics[width=\linewidth]{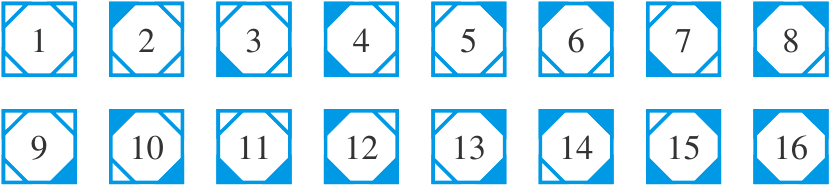}
	\caption{The complete set of corner Wang tiles over two colors.}
	\label{fig:complete_set}
\end{figure}

The corner Wang tiles proved to be preferable over the traditional Wang tiles, allowing for simpler generation of valid tilings, reduced memory requirements, and easier generalization to multiple dimensions~\citep{Lagae:2006:AWT}, while preserving the possibility of building aperiodic tilings~\citep{Lagae:2006:ASS}. These findings inspired us to employ corner tiles in this paper. In particular, we consider here the complete set of planar corner Wang tiles over two colors, containing one corner tile for each possible combination of color codes as depicted in Fig.~\ref{fig:complete_set}.

\begin{figure}[t]
	\includegraphics[width=\linewidth]{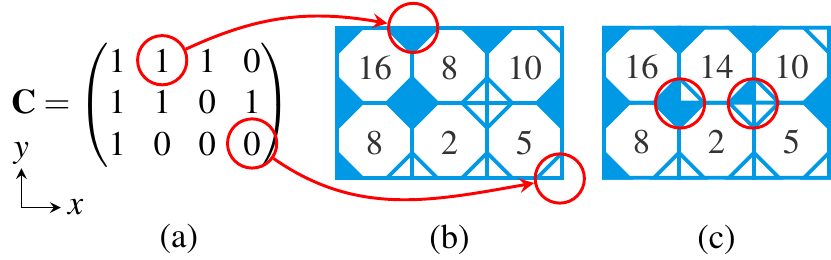}
	\caption{Illustration of (a) a connectivity matrix and (b) its correspondence to a~valid tiling, (c) an example of an invalid tiling.}
	\label{fig:tiling}
\end{figure}

Using corner tiles, valid assemblies or tilings must satisfy identical colorings of shared vertices over all adjacent tiles, compare Fig. \ref{fig:tiling}b and \ref{fig:tiling}c. Marking each color code with an integer value~\citep{Lagae:2006:AWT}, any valid assembly determines the color-code connectivity matrix $\mathbf{C}$ uniquely. Note that in the case of only two vertex colors, the connectivity matrix becomes Boolean, Fig.~\ref{fig:tiling}a.
\nomenclature{$\mathbf{C}$}{Connectivity matrix} %
\nomenclature{$n_{\mathrm{t,x}}$}{Number of tiles in horizontal direction} %
\nomenclature{$n_{\mathrm{t,y}}$}{Number of tiles in vertical direction} %
Conversely, for all complete sets of corner tiles over a~limited set of colors, any connectivity matrix containing integer values corresponding to the vertex codes of the set automatically defines a~valid rectangular tiling. Notice that an extension to non-rectangular tiling with holes is straightforward, using a~flattened one-dimensional array.

\subsection{Optimal truss design}

Trusses are structures consisting of nodes and straight bars which transmit axial forces only. While the optimal topology of the least-compliant trusses under a~single load case aligns structural stiffness with the principal strains~\citep{Michell:1904:LEM}, their trajectories are not straight in general and hence an optimal design can contain an infinite number of bars. To overcome this undesirable property, the continuum design domain is usually discretized into the so-called ground-structure~\citep{Dorn:1964:ADO}, constituting a~finite-dimensional design space formed by the sets of $n_\mathrm{n} \in \mathbb{N}$ fixed nodes and $n_\mathrm{b} \in \mathbb{N}$ potential bars. Design variables of truss topology optimization then involve cross-section areas $\mathbf{a} \in \mathbb{R}^{n_\mathrm{b}}_{\ge 0}$~\citep{Bendsoe:2003:TO}, possibly attaining zero values, so that the structurally inefficient truss elements vanish. To this goal, two branches of truss topology optimization commonly apply. 
\nomenclature{$\mathcal{D}$}{Fixed design domain}
\nomenclature{$n_\mathrm{n}$}{Number of nodes}%
\nomenclature{$n_\mathrm{b}$}{Number of bars}%
\nomenclature{$\mathbf{a}$}{Cross-section areas column vector}%

The traditional \textit{plastic design}~\citep{Dorn:1964:ADO} poses the problem in terms of the member axial forces only, and searches for the minimum-weight topology under the static equilibrium constraint and bounds on allowed stresses. Despite neglecting the kinemat\-ic compatibility, convergence to statically determinate\footnote{Statically determinate designs uniquely determine the axial forces only based on the static equilibrium~\citep{Rozvany:2014:FEM}.} optimal designs occurs~\citep{Sved:195:MWC} when a~single load case is considered~\citep{Rozvany:2014:FEM} and cross-section areas are not constrained, so that the compatibility conditions hold. Unfortunately, modularity violates the latter assumption by implicitly requiring equality of cross-sections within specified sets of elements. This can be seen, for example, by optimizing any statically indeterminate ground structure and requiring all bars to share the same cross-section area. Consequently, modular designs obtained by the plastic formulation may and usually are statically indeterminate, violate the compatibility conditions, and in turn provide a~lower bound for the objective function value only.

Therefore, the compatibility conditions must be considered, impelling us to use an \textit{elastic design} formulation instead. Here, we search for the least-compliant design \eqref{eq:compliance}, subjected to the bound on the available volume of material \eqref{eq:volume} and the equilibrium equation \eqref{eq:compatibility}, i.e., 
\begin{subequations}\label{eq:trusstopo_nonlin}
\begin{align}
	\underset{\mathbf{a}, \mathbf{u}}{\min}\quad & \frac{1}{2}\mathbf{f}^\mathrm{T} \mathbf{u}\label{eq:compliance}\\
	\mathrm{s.t.}\quad & \mathbf{K}(\mathbf{a}) \mathbf{u} = \mathbf{f},\label{eq:compatibility}\\
	              & \bm{\ell}^\mathrm{T} \mathbf{a} \le \overline{V},\label{eq:volume}\\
	              & \mathbf{a} \ge \mathbf{0},
\end{align}
\end{subequations}
\nomenclature{$\mathbf{K}$}{Stiffness matrix}%
\nomenclature{$\mathbf{u}$}{Displacements column vector}%
\nomenclature{$\mathbf{f}$}{Nodal forces column vector}%
\nomenclature{$\bm{\ell}$}{Bars' lengths column vector}%
\nomenclature{$\overline{V}$}{Volume upper-bound}%
\nomenclature{$\overline{n_{\mathrm{dof}}}$}{Number of degrees of freedom}%
where, $\mathbf{f} \in \mathbb{R}^{n_\mathrm{dof}}$ denotes the nodal forces column vector, $n_\mathrm{dof} \in \mathbb{N}$ refers to the number of degrees of freedom, $\mathbf{u} \in \mathbb{R}^{n_\mathrm{dof}}$ stands for the displacements column vector, $\mathbf{K(\mathbf{a})} \in \mathbb{R}^{n_\mathrm{dof} \times n_\mathrm{dof}}$ is the structural stiffness matrix, an affine function of the cross-section areas, $\bm{\ell} \in \mathbb{R}^{n_\mathrm{b}}$ stands for the column vector of the bars lengths, and $\overline{V} \in \mathbb{R}_{>0}$ denotes an upper-bound on the total structural volume. Finally, the objective of the optimization $c = \frac{1}{2} \mathbf{f}^\mathrm{T} \mathbf{u}$, $c \in \mathbb{R}_{\ge 0}$, \eqref{eq:compliance} denotes compliance, i.e., half of the work done by the external loads.
\nomenclature{$c$}{Compliance}

While the problem \eqref{eq:trusstopo_nonlin} is straightforward to formulate, it lacks convexity due to the bilinear equilibrium equation \eqref{eq:compatibility} with a~possibly singular stiffness matrix, and is hard to solve (to global optimality) even for small-scale problems~\citep{Kocvara:2006:ERT}. However, based on~\citep{Lobo:1998:ASO} and Section 3.4.3 in~\citep{Ben-Tal:2001:LMC}, we can write (its dual) convex second-order cone programming (SOCP) reformulation:
\begin{subequations}\label{eq:trusstopo_socp}
	\begin{align}
	\min_{%
		\mathbf{a},
		\mathbf{w},
		\mathbf{s}
		} 
	\quad & 
	\mathbf{1}^\mathrm{T} \mathbf{w} \label{eq:trusstopo_socp_compl}\\
	\mathrm{s.t.} \quad 
	& \bm{\ell}^\mathrm{T} \mathbf{a} \le \overline{V},\label{eq:trusstopo_socp_volume}\\
	& 
	\mathbf{A} \mathbf{s} = \mathbf{f},\label{eq:trusstopo_socp_static}\\
	& \left\lVert
	\begin{pmatrix}
	w_i - a_i\\
	\sqrt{\frac{2 \ell_i}{E_i}} s_i
	\end{pmatrix} 
	\right\rVert_2 \le w_i + a_i, 
	\quad\forall i \in \{1\;..\;n_\mathrm{b} \},\label{eq:trusstopo_socp_socp}\\
	& \mathbf{a} \ge \mathbf{0}\label{eq:trusstopo_socp_areas},
	\end{align}
\end{subequations}
which is efficiently solvable to global optimality by interior-point methods~\citep{Anjos:2012:HSC}. In Eq. \eqref{eq:trusstopo_socp}, the symbol $\mathbf{s} \in \mathbb{R}^{n_\mathrm{b}}$ stands for the axial forces column vector, $E_i \in \mathbb{R}_{> 0}$ denotes the modulus of elasticity, $w_i \in \mathbb{R}_{\ge0}$ constitutes the complementary strain energy, and $\ell_i$ is the length of the $i$-th bar, respectively. Further, $\mathbf{A} \in \mathbb{R}^{n_\mathrm{dof}\times n_\mathrm{b}}$ stands for the static matrix relating axial, $\mathbf{s}$, and nodal, $\mathbf{f}$, forces. At the optimum, the objective function \eqref{eq:trusstopo_socp_compl} attains the value of the complementary strain energy, equal to the compliance, as $\mathbf{w}$ provides a~to-be-minimized upper bounds on the complementary strain energies of individual bars in the second-order conic constraints \eqref{eq:trusstopo_socp_socp}.

\section{Methodology}\label{sec:methodology}

Modularity constitutes a~partitioning of a~complex structure into several simpler repeated units---modules. Here, we assume that a~fixed (rectangular) structural design domain of the size $n_{\mathrm{t},y} \times n_{\mathrm{t},x}$ consists of square truss modules with a~fixed orientation, see Fig.~\ref{fig:beam_modular}a. Without loss of generality, the number of (employed) module types $n_\mathrm{t}$ is at most $n_{\mathrm{t},y} n_{\mathrm{t},x}$. With $n_\mathrm{t} = n_{\mathrm{t},y} n_{\mathrm{t},x}$, the problem is equivalent to the non-modular design, while $n_{\mathrm{t}} = 1$ implies a~single-module periodic design.

When $n_t \ll n_{\mathrm{t},y} n_{\mathrm{t},x}$, it may happen that each module type neighbors with all the remaining module types, implying $n_t^2$ ways of possible module interconnections. In this setting, it is therefore not surprising that a~single solid/high-stiffness interface is obtained during optimization~\citep{Garner2019} and this interface then propagates periodically through the macro design domain. Aiming at controlling the numbers of modules and their interface types directly, Wang tiles appear to be a~natural approach. 

Therefore, we further restrict our formalism of describing modular assembly plans to Wang tilings. For the sake of demonstration, we consider here the complete set of corner Wang tiles over two colors, recall Section \ref{sec:vertex_based}. Consequently, we have $n_t = 16$ together with four independent horizontal and vertical edge types. Discretized by trusses, these modules are compatible by definition over the matching edges in the sense of generating a~statically admissible ground structure.

In this section, we first suitably modify the SOCP formulation for truss topology optimization \eqref{eq:trusstopo_socp} to account for structural modularity and to handle multiple loading conditions, stress constraints, and module reusability. Ultimately, we develop a~bilevel optimization approach to optimize the topologies of all the module types and their assemblies simultaneously.
\nomenclature{$n_\mathrm{t}$}{Number of module types}

\subsection{Truss topology optimization extended to structural modularity}

Structural modularity is inherently prescribed in the form of equality constraints of certain cross-section areas, consequently reducing the number of \textit{unique} cross-section areas within the ground structure. The developed formulation preserves its convexity and applies when a~fixed module assembly plan $\mathbf{C}$ is specified a~priori. 

\begin{figure}[b]
	\centering
	\includegraphics[width=\linewidth]{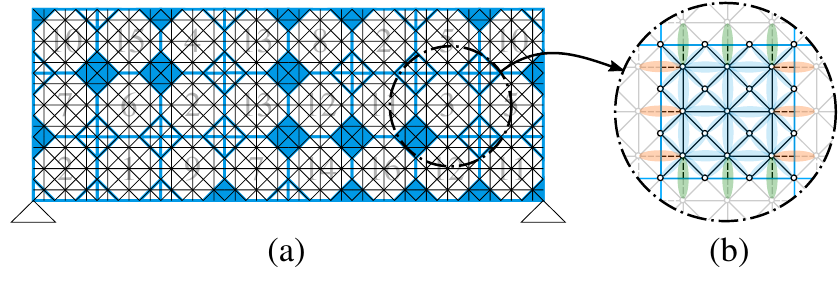}
	\caption{(a) Partitioning of a~design domain into modules. (b) Bars forming each module split up into two sets---module-associated (highlighted in blue), and edge-associated (shown with green and red backgrounds).}
	\label{fig:beam_modular}
\end{figure}

In the following text, we consider that all truss modules share the same \textit{module ground structure} inspired by the union-jack lattice \citep{Gurtner2014}, see Fig. \ref{fig:beam_modular}. Another ground structure can be adopted though, if needed. 

To secure the mechanical compatibility of modules over their edges, the bars within each module split up into two sets. Those located entirely inside a~module---the \textit{module-associated} bars---occur only in the specific module type, and are highlighted in blue in Fig. \ref{fig:beam_modular}b. The second set contains bars that intersect module boundaries. To preserve constant cross-section areas of the inter-domain bars along their lengths, they need to be associated with the edge types, occurring in multiple modules. These \textit{edge-associated} bars are highlighted in red and green in Fig. \ref{fig:beam_modular}b. Distinction of horizontal (green) and vertical (red) bars is the consequence of forbidden rotations in Wang tiling formalism.

\begin{figure}[t]
	\centering
	\includegraphics[width=\linewidth]{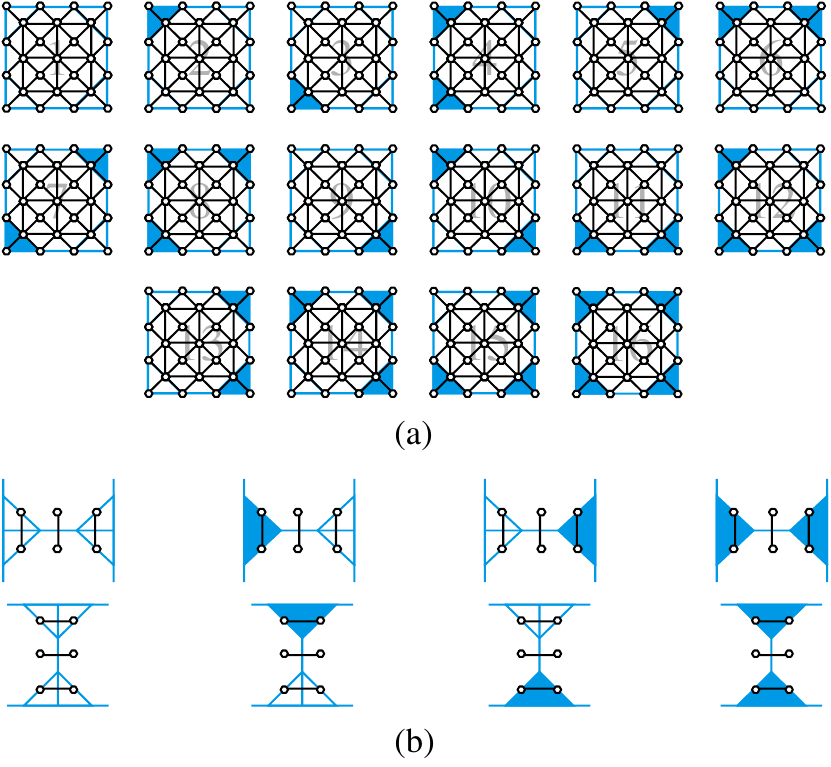}
	\caption{(a) Module- and (b) edge-associated bars. Scattered points represent nodes.}
	\label{fig:tile-associated}
\end{figure}

To achieve identical topology of the modules and their edge connections in all their occurrences, the bars are divided into groups, such that all bars in the same group share the same cross-section area. Assignment of bars to a particular group is provided by the group vector $\mathbf{g} (\mathbf{C}) \in \mathbb{N}^{n_\mathrm{b}}$, uniquely for each assembly plan $\mathbf{C}$. This group vector assigns a~single number in the range of $\{ 1 \;..\; n_\mathrm{g} \}$ to each bar of the ground structure, where $n_\mathrm{g}$ denotes the number of groups, i.e., the number of unique cross-section areas. 

For the complete tileset shown in Fig. \ref{fig:complete_set} and the module ground structure, Fig. \ref{fig:beam_modular}b, we have $48$ bars associated with each module type, i.e., $16\times48 = 768$ groups of module-associated bars in total, see Fig. \ref{fig:tile-associated}a. Similarly, each edge, either horizontal or vertical, accommodates $3$ edge-associated bars, leading to $8\times3 = 24$ groups of edge-associated bars, see Fig. \ref{fig:tile-associated}b. Consequently, we have $n_\mathrm{g} = 792$ for this specific choice of the tileset and the module ground structure.
\nomenclature{$n_\mathrm{g}$}{Number of groups}%
\nomenclature{$\mathbf{a}_\mathrm{G}$}{Group matrix}%

Following the definition of the group vector, let $\mathbf{G} \in \mathbb{B}^{n_\mathrm{b} \times n_\mathrm{g}}$ denote the group matrix defined as
\begin{equation}\label{eq:group_matrix}
	G_{i,j} (\mathbf{C}) = 
	\begin{dcases}
		0\quad\text{if } j \neq g_i (\mathbf{C}),\\
		1\quad\text{if } j  = g_i(\mathbf{C}),
	\end{dcases} \quad \forall i \in \{ 1 \;..\; n_\mathrm{b} \}, \forall j \in \{ 1 \;..\; n_\mathrm{g} \},
\end{equation}
with $G_{i,j} (\mathbf{C})$ being the element in the $i$-th row and $j$-th column of the group matrix $\mathbf{G} (\mathbf{C})$, and $g_i$ standing for the $i$-th element in $\mathbf{g}$. The group matrix represents a~linear transformation mapping the space of the unique cross-section areas $\mathbf{a}_\mathrm{g} \in \mathbb{R}^{n_\mathrm{g}}$ into the space of all cross-section areas $\mathbf{a} \in \mathbb{R}^{n_\mathrm{b}}$,
\begin{equation}
	\mathbf{a} = \mathbf{G}(\mathbf{C}) \mathbf{a}_\mathrm{g}.
\end{equation}

Because of modularity, the original topology optimization formulation \eqref{eq:trusstopo_socp} must be modified, reducing the number of cross-section areas from $n_{\mathrm{b}}$ to $n_{\mathrm{g}}$, as they are substituted by unique cross-section areas\footnote{Note that if $n_\mathrm{g} = n_\mathrm{b}$, the problem simplifies back to non-modular design.}. Moreover, a~similar procedure can be applied to reduce the number of the complementary strain energies of bars $\mathbf{w} \in \mathbb{R}^{n_\mathrm{b}}$ to $\mathbf{w}_\mathrm{g} \in \mathbb{R}^{n_\mathrm{g}}$, where $w_{\mathrm{g},j}$ is the complementary strain energy of all the bars in group $j$.
\nomenclature{$\textbf{a}_\mathrm{g}$}{Grouped cross-section areas column vector}%
\nomenclature{$\textbf{w}_\mathrm{g}$}{Grouped complementary strain energy column vector}%
Consequently, the objective function reads as
\begin{equation}\label{eq:obj_groups}
\mathbf{1}^\mathrm{T} \textbf{w}_{\mathrm{g}},
\end{equation}
and the volume constraint \eqref{eq:trusstopo_socp_volume} transforms into
\begin{equation}\label{eq:volume_groups}
	\bm{\ell}^{\mathrm{T}} \textbf{G}(\mathbf{C}) \textbf{a}_\mathrm{g} \le \overline{V}.
\end{equation}
For the second-order conic constraints \eqref{eq:trusstopo_socp_socp}, we follow the aggregation in Appendix~\ref{app:SOCP} to receive
\begin{equation}
\begin{multlined}
w_{\mathrm{g},j} + a_{\mathrm{g},j} \ge \left\lVert 
\begin{pmatrix}
w_{\mathrm{g},j} - a_{\mathrm{g},j}\\
\textbf{G}_{:,j}(\mathbf{C}) \odot \left[2 \bm{\ell} \oslash \mathbf{E}
\right]^{\circ\frac{1}{2}} \odot \mathbf{s} 
\end{pmatrix}
\right\rVert_2,\\
\forall j \in \{1\;..\;n_\mathrm{g}\},
\end{multlined}
\end{equation}
with $\oslash$, $\odot$, and $^{\circ}$ denoting the Hadamard (element-wise) division, multiplication, and power; and $\textbf{G}_{:,j}$ being the $j$-th column of $\mathbf{G}$.

The final formulation of truss topology optimization extended to structural modularity then reads
\begin{subequations}\label{eq:trusstopo_socpmod}
	\begin{align}
	\underset{\mathbf{a}_\mathrm{g}, \mathbf{w}_\mathrm{g}, \mathbf{s}}{ \min}\; & \mathbf{1}^\mathrm{T} \mathbf{w}_\mathrm{g} \label{eq:trusstopo_socpmod_compl}\\
	\mathrm{s.t.}\quad & \bm{\ell}^{\mathrm{T}} \textbf{G}(\mathbf{C}) \textbf{a}_\mathrm{g} \le \overline{V},\label{eq:trusstopo_socpmod_volume}\\
	& \mathbf{A}\mathbf{s} = \mathbf{f},\label{eq:trusstopo_socpmod_static}\\
	& \begin{aligned} 
	w_{\mathrm{g},j} + a_{\mathrm{g},j} \ge \left\lVert 
	\begin{pmatrix}
	w_{\mathrm{g},j} - a_{\mathrm{g},j}\\
	\textbf{G}_{:,j}(\mathbf{C}) \odot \left[2 \bm{\ell} \oslash \mathbf{E}
	\right]^{\circ\frac{1}{2}} \odot \mathbf{s}
	\end{pmatrix}
	\right\rVert_2,\\ \forall j \in \{1\;..\;n_\mathrm{g}\},
	\end{aligned}\label{eq:trusstopo_socpmod_socp}\\
	& \mathbf{a}_\mathrm{g} \ge \mathbf{0}.
	\end{align}
\end{subequations}
Because the number of constraints and variables in \eqref{eq:trusstopo_socpmod} decreases compared to the non-modular design \eqref{eq:trusstopo_socp}, finding the optimal topology of a~modular structure is faster than obtaining the optimal non-modular design. The acceleration factor depends on the repeatability of individual module types and, of course, on the optimization solver employed.

\begin{figure*}[!b]
	\normalsize
	\setcounter{equation}{10}
	\begin{subequations}\label{eq:trusstopo_socpmod2}
		\begin{align}
		\begin{pmatrix}
		\mathbf{a}_\mathrm{g}^* \\ \mathbf{w}_\mathrm{g}^* \\ \mathbf{s}^*
		\end{pmatrix}
		\in 
		\underset{\mathbf{a}_\mathrm{g}, \mathbf{w}_\mathrm{g}, \mathbf{s}}{\min}\quad & \mathbf{1}^\mathrm{T} \mathbf{w}_\mathrm{g}\\
		\mathrm{s.t.}\quad & \sum_{m=1}^{n_\mathrm{s}} \left(\bm{\ell}^{(m)}\right)^{\mathrm{T}} \left[\textbf{G}(\mathbf{C})\right]^{(m)} \textbf{a}_\mathrm{g} \le \overline{V},\\
		& \mathbf{A}\mathbf{s}_{k}^{(m)} = \mathbf{f}_{k}^{(m)},& \forall k \in \{1\;..\;n_\mathrm{lc}\}, \forall m \in \{1\;..\;n_\mathrm{s}\},\\	
		& w_{\mathrm{g},j} + a_{\mathrm{g},j} \ge \left\lVert 
		\begin{pmatrix}
		w_{\mathrm{g},j} - a_{\mathrm{g},j}\\
		\omega_{1,1} \left[\textbf{G}_{:,j}(\mathbf{C})\right]^{(1)} \odot \left[2 \bm{\ell^{(1)}} \oslash \mathbf{E}
		\right]^{\circ\frac{1}{2}} \odot \mathbf{s}_{1}^{(1)}\\
		\vdots\\
		\omega_{n_\mathrm{lc},1} \left[\textbf{G}_{:,j}(\mathbf{C})\right]^{(1)} \odot \left[2 \bm{\ell}^{(1)} \oslash \mathbf{E}\right]^{\circ\frac{1}{2}} \odot \mathbf{s}_{{n_\mathrm{lc}}}^{(1)}\\
		\omega_{1,2} \left[\textbf{G}_{:,j}(\mathbf{C})\right]^{(2)} \odot \left[2 \bm{\ell}^{(2)} \oslash \mathbf{E}\right]^{\circ\frac{1}{2}} \odot \mathbf{s}_{{1}}^{(2)}\\
		\vdots\\
		\omega_{n_\mathrm{lc},n_\mathrm{s}} \left[\textbf{G}_{:,j}(\mathbf{C})\right]^{(n_\mathrm{s})} \odot \left[2 \bm{\ell}^{(n_\mathrm{s})} \oslash \mathbf{E}\right]^{\circ\frac{1}{2}} \odot \mathbf{s}_{{n_\mathrm{lc}}}^{(n_\mathrm{s})}
		\end{pmatrix}
		\right\rVert_2,&\forall j \in \{1\;..\;n_\mathrm{g}\},\\
		& \sigma_\mathrm{L} \left[\textbf{G}(\mathbf{C})\right]^{(m)} \textbf{a}_\mathrm{g} \le \textbf{s}_{k}^{(m)} \le \sigma_\mathrm{U} \left[\textbf{G}(\mathbf{C})\right]^{(m)} \textbf{a}_\mathrm{g}, &\forall k \in \{1\;..\;n_\mathrm{lc}\}, \forall m \in \{1\;..\;n_\mathrm{s}\}, \\
		& \mathbf{a}_\mathrm{g} \ge \mathbf{0}.
		\end{align}%
	\end{subequations}%
\end{figure*}

\subsection{Handling stress constraints, multiple load cases, and module reusability}
In addition to modularity, we extend formulation \eqref{eq:trusstopo_socpmod} to handle stress constraints, multiple loading scenarios, and module reusability among multiple structures. Most importantly, all these extensions preserve the convexity of the optimization problem, and thus do not compromise the efficiency of the solution.

In the case of stress constraints, let us assume that
\begin{equation}\label{eq:stress}
\setcounter{equation}{9}
\sigma_\mathrm{L} \le \sigma_i \le \sigma_\mathrm{U}, \quad\forall i \in \{1\;..\;n_\mathrm{b}\},
\end{equation}%
where $\sigma_\mathrm{L}$ and $\sigma_\mathrm{U}$ are the bounds for stress in element $i$, $\sigma_i$. To maintain convexity and avoid additional variables, we can multiply the inequality \eqref{eq:stress} by $a_i$ and constrain the internal forces, $s_i$, instead of the stress variables, $\sigma_i$, which provides us with equivalent convex linear inequalities
\begin{equation}
\setcounter{equation}{10}
\sigma_\mathrm{L} \textbf{G}(\mathbf{C}) \textbf{a}_\mathrm{g} \le \textbf{s} \le \sigma_\mathrm{U} \textbf{G}(\mathbf{C}) \textbf{a}_\mathrm{g}.
\end{equation}
\setcounter{equation}{11}%
For multiple loading conditions and module reusability, we assume that there are $n_\mathrm{s} \in \mathbb{N}$ structures subjected to $n_\mathrm{lc} \in \mathbb{N}$ load cases, and we minimize the weighted average of the complementary strain energies of all the load cases with the weights $\bm{\omega} \in \mathbb{R}_{>0}^{n_\mathrm{lc} \times n_\mathrm{s}}$. To minimize the number of design variables and constraints, we aggregate the complementary strain energies of bars with the same cross-sections across load cases. Because this aggregation follows the steps already outlined in Appendix~\ref{app:SOCP}, we omit the details here for the sake of brevity.

The final convex second-order conic formulation for optimization of trusses with prescribed modularity and subjected to stress constraints, multiple loading scenarios, and allowing for module reusability reads as \eqref{eq:trusstopo_socpmod2}, where $\mathbf{s}_k$ and $\mathbf{f}_k$ stand for the axial and nodal forces associated with the $k$-th load case. Moreover, the superscript $\bullet^{(m)}$ associates the variable $\bullet$ with the $m$-th structure.

\subsection{Modular-topology optimization}

The objective function of the optimal design obtained by solving \eqref{eq:trusstopo_socpmod} or \eqref{eq:trusstopo_socpmod2} depends inherently on the specified assembly plan of modules $\mathbf{C}$. However, because the number of potential valid assemblies increases with the number of entries in $\mathbf{C}$ exponentially, exploring all possible combinations may be untractable. Therefore, a~method to efficiently find a~“good” connectivity matrix must be developed. In this section, we propose an approach to solve this bilevel optimization problem, i.e., optimizing the module topologies as well as their assembly simultaneously. While the lower-level problem \eqref{eq:trusstopo_socpmod2} exhibits convexity, the upper-level is combinatorial and NP-hard in general~\citep{Demaine2007}. Therefore, we propose tackling the problem with a~combination of mathematical programming and meta-heuristics.

The bilevel optimization problem then reads
\begin{align}\label{eq:MTO_SOCP}
\mathbf{C}^* 
\in
\argmin_{\mathbf{C}}\;
\mathbf{1}^\mathrm{T} \mathbf{w}_{\mathrm{g}}^*( \mathbf{C} ),
\end{align}
with $w^*$ following from \eqref{eq:trusstopo_socpmod2}, and the globally optimal design is eventually recovered as
\begin{align}
\mathbf{a}^* = \mathbf{G}(\mathbf{C}^*) \mathbf{a}_\mathrm{g}^*(\mathbf{C}^*).
\end{align}
The problem \eqref{eq:MTO_SOCP} is solved with (i) the globally optimal connectivity matrix $\mathbf{C}^*$ and with (ii) the globally optimal vector of unique cross-section areas $\mathbf{a}_\mathrm{g}^*$ at the globally optimal complementary strain energy $c^*$. 

To approximately solve the upper-level assembly problem, we adopt the genetic algorithm (GA)~\citep{Holland:1992:ANA}, a~stochastic meta-heuristic optimization algorithm that simulates the process of evolution by following Darwin’s “survival of the fittest” rule~\citep{Spencer:1864:PB}. Genetic algorithms receive high recognition in combinatorial and multi-objective optimizations, as they handle discrete, non-differentiable, and non-convex optimization problems fairly efficiently. Here, we implement the standard GA routine \citep{Holland:1992:ANA}, see Algorithm \ref{alg:GA}, consisting of the following steps: (i) First, we generate a~random population of $n_\mathrm{pop}$ individuals (connectivity matrices) that evolve through $n_\mathrm{gen}$ generations. (ii) Fitness of individuals, inversely proportional to the complementary strain energy, result from a~parallel solution to \eqref{eq:trusstopo_socpmod2}. Next, (iii) tournament selection is performed, controlled by probability $p_\mathrm{t}$ and the number of competitors $n_\mathrm{t}$. Both the (iv) cross-over and (v) mutation operators are applied, governed by the probabilities $p_\mathrm{c}$ and $p_\mathrm{m}$, and following the standard implementations. Finally, we also enforce population diversity, so that duplicate individuals get substituted by random ones, and elitism applies. Specific numerical values of these parameters appear in Appendix~\ref{app:GA}.

\begin{algorithm}[t].
	\caption{Genetic Algorithm}
	\begin{algorithmic}
		\Function{\texttt{GeneticAlgorithm}}{$\mathbf{C}$, $n_\mathrm{pop}$, $n_\mathrm{gen}$, $n_\mathrm{t}$, $p_\mathrm{t}$, $p_\mathrm{c}$, $p_\mathrm{m}$}
		\State $population \gets \texttt{RandomPopulation}(\mathbf{C}, n_\mathrm{pop})$
		\State $fitness \gets \texttt{PopulationFitness}(population)$
		\For{$\delta \gets \{1\;..\; n_\mathrm{gen}\}$}
		\State $elite \gets \texttt{EliteIndividual}(population,fitness)$
		\State $matingPool \gets \texttt{Selection}(population, fitness, n_\mathrm{t}, p_\mathrm{t})$
		\State $population \gets \texttt{CrossOver}(matingPool, p_\mathrm{c})$
		\State $population \gets \texttt{Mutation}(population, p_\mathrm{m})$
		\State $population \gets \texttt{AppendElite}(population, elite)$
		\State $population \gets \texttt{Diversify}(population)$
		\State $fitness \gets \texttt{PopulationFitness}(population)$
		\EndFor
		\State \Return $\mathbf{C} \gets \texttt{EliteIndividual}(population)$
		\EndFunction
	\end{algorithmic}
	\label{alg:GA}
\end{algorithm}

\section{Examples}\label{sec:example}

The proposed modular-topology optimization approach was implemented in \textsc{Matlab}, with the source codes available at \citep{tyburec_marek_2020_3750751}, and applied to three illustrative two-dimensional problems. In Section \ref{sec:hinge}, we consider a~hinge-supported beam subjected to a~single load case without stress constraints. In this setting, we first adopt coarse discretization, as this allows us to obtain a~globally optimal design through brute-force enumeration. Subsequently, we assume a~finer discretization and discuss the results. Section \ref{sec:L} investigates an L-shaped domain with two load cases and stress constraints. The third problem in Section \ref{sec:reusability} aims at designing modules and assemblies that are reusable for both of the former cases.

All computations were performed on a~Linux workstation with two Intel\textsuperscript{\textregistered} Xeon\textsuperscript{\textregistered} E5-2630 processors. The second-order cone programs for topology optimization of trusses were solved using the state-of-the-art \textsc{Mosek} optimizer~\citep{mosek}, interfaced with \textsc{Matlab} via the \textsc{Yalmip} toolbox~\citep{Lofberg:2004:YTM}.

\subsection{Hinge-supported beam}\label{sec:hinge}
\subsubsection{Coarse discretization}

\begin{figure}[t]
	\centering
	\includegraphics[width=\linewidth]{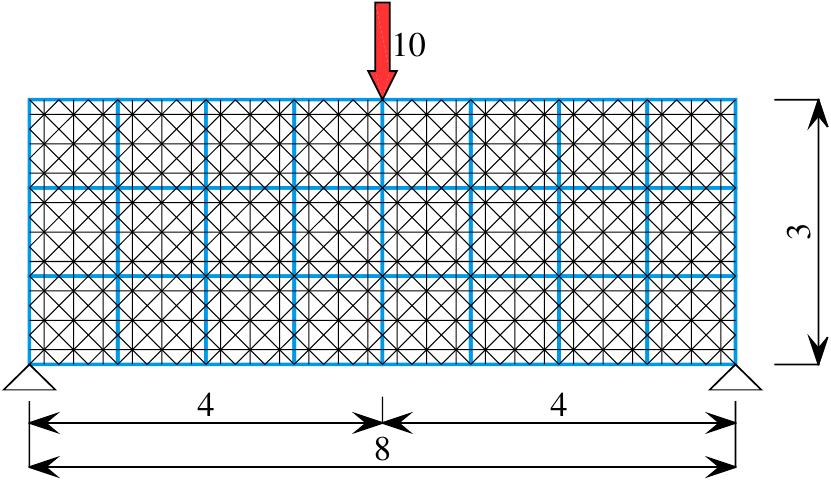}
	\caption{Dimensions, discretization into modules, boundary conditions, and ground structure of the coarsely discretized beam.}
	\label{fig:beam_dimensions}
\end{figure}

As the first illustration, we investigate a~simply-supported beam of dimensions $8 \times 3$, see Fig. \ref{fig:beam_dimensions}. Under coarse discretization, the beam splits up into $24$ unit-size square modules that follow the corner Wang tiling formalism, recall Fig.~\ref{fig:complete_set}. We assume the module ground structures shown in Fig. \ref{fig:beam_modular}b, with Young's modulus $E$ of each bar equal to $1$. The beam is supported with two hinges at the very bottom-left and bottom-right corners, and loaded with an external force of magnitude $10$ at the mid-span of the top edge.

\begin{figure}[t]
	\centering
	\begin{subfigure}{\linewidth}
		\includegraphics[width=8.4cm]{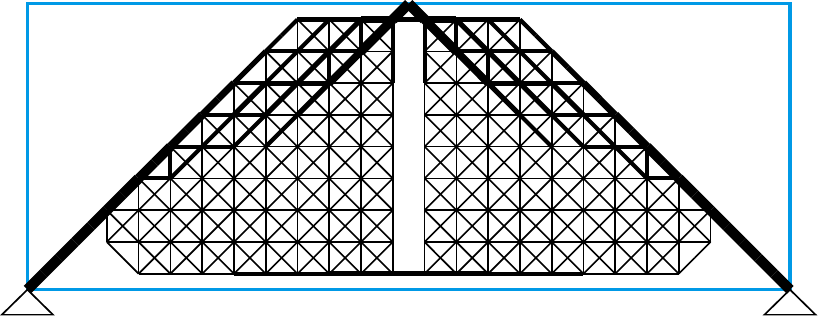}
		\caption{}
		\label{fig:beam_ideal}
	\end{subfigure}\\
	\begin{subfigure}{\linewidth}
		\includegraphics[width=8.4cm]{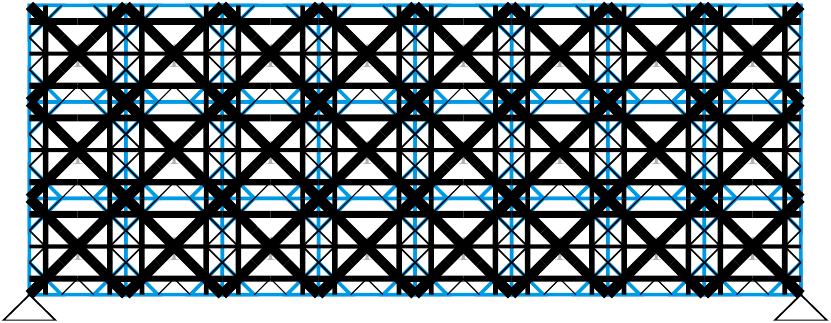}
		\caption{}
		\label{fig:beam_worst}
	\end{subfigure}
	\caption{(a) Lower-bound non-modular, and (b) worst-case designs of the evaluated coarsely discretized beam with complementary strain energies $\underline{c}=61.9$ and $\overline{c}=191.2$, respectively.}
\end{figure}

\begin{figure}[b]
	\setcounter{figure}{6}
	\centering
	\begin{subfigure}{\linewidth}
		\includegraphics[width=\linewidth]{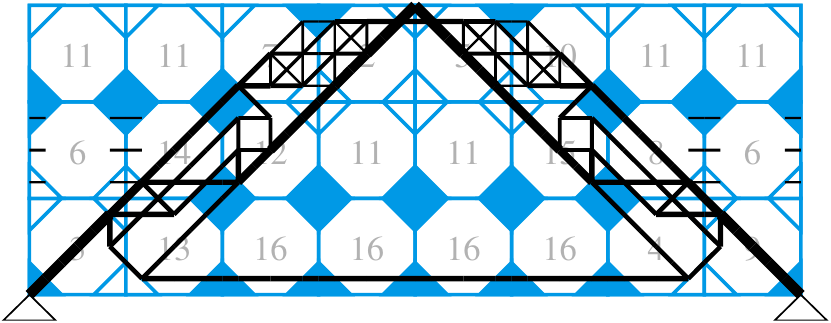}
		\caption{}
		\label{fig:beam_global}
	\end{subfigure}\\
	\begin{subfigure}{\linewidth}
		\includegraphics[width=\linewidth]{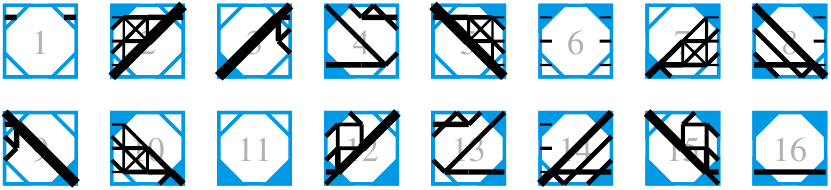}
		\caption{}
		\label{fig:beam_globaltileset}
	\end{subfigure}
	\caption{Globally optimal design for the (a) coarsely discretized beam of $c^* = 62.7$, and (b) the corresponding tileset.}
\end{figure}

\begin{figure}[!t]
	\setcounter{figure}{7}
	\centering
	\def\svgwidth{84mm}
	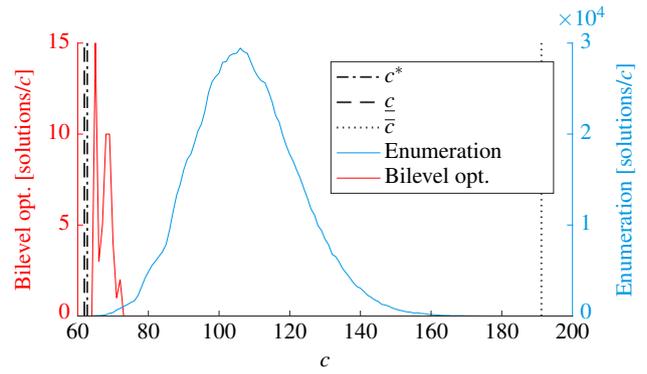
	\caption{Distribution of optimal complementary strain energies $c$ of all enumerated combinations (in blue), and $50$ independent runs of the bilevel optimization (in red). While $\overline{c}$ and $c^*$ denote the complementary strain energies of the worst and best modular designs, $\underline{c}$ is the complementary strain energy of the non-modular one.}
	\label{fig:histogram}
\end{figure}

\begin{figure*}[!t]
	\setcounter{figure}{8}
	\centering
	\includegraphics[width=\linewidth]{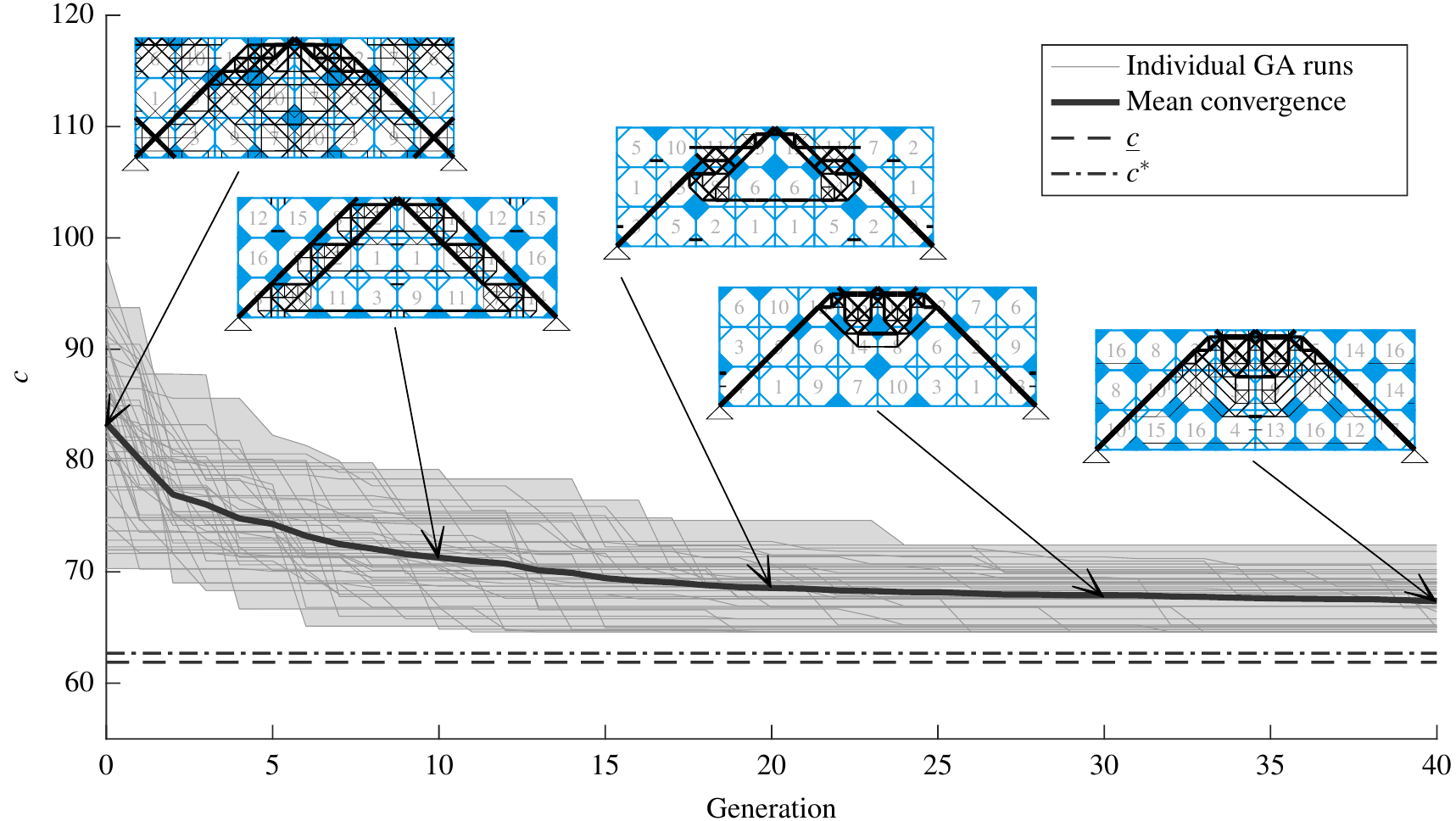}
	\caption{Convergence of complementary strain energies $c$ of the best individuals with $50$ independent runs of the genetic algorithm; $c^*$ denotes the complementary strain energies for the best modular design and $\underline{c}$ stands for the complementary strain energy of the non-modular one.}
	\label{fig:GA}
\end{figure*}

\paragraph{Bounds on the global optimum}
Structural modularity comes at a~price of higher complementary strain energy when compared to the non-modular design~\citep{Huang:2008:ODP}. Because all the modules share identical module ground structures, performing topology optimization without modularity constraints, recall Eq. \eqref{eq:trusstopo_socp}, provides the lower-bound energy $\underline{c} = 61.9$ with the design shown in Fig. \ref{fig:beam_ideal}. Analogously, the upper-bound complementary strain energy arises in the topology optimization of the design domain assembled from a~single module type, indicated by the connectivity matrix $\textbf{C}$ containing all-zeros or all-ones. In this setting, topology optimization results in the optimal design depicted in Fig. \ref{fig:beam_worst}, with $\overline{c} = 191.2$. Therefore, the strain energy of the optimal modular design must lie in the interval $c^* \in \left[ 61.9, 191.2 \right]$.

\paragraph{Complete enumeration}

Thanks to coarse discretization, the globally optimal connectivity matrix $\mathbf{C}^*$ and the corresponding complementary strain energy $c^*$ can be determined using the complete enumeration, which allows us to assess the performance of the bilevel optimization rigorously. Notice that even the coarse discretization permits $2^{36}$ possible connectivity matrices, preventing a~complete enumeration in a~reasonable time. To reduce the complexity, we have enforced symmetry of the connectivity matrix $\textbf{C}$ with respect to the vertical axis of the beam, consequently reducing the number of combinations to $2^{20}$ feasible assemblies. Because all modules comprise identical module ground structures, the number of combinations is further decreased by noticing that the vertex types lack any physical meaning. This makes the problem invariant against the coloring of Wang tiles corners, i.e., $\mathbf{w}_\mathrm{g}(\mathbf{C}) = \mathbf{w}_\mathrm{g}(\overline{\mathbf{C}})$, where the connectivity matrix $\overline{\mathbf{C}}$ follows from $\mathbf{C}$ by inverting $0$ to $1$ and vice versa. Subsequently, we need to enumerate $2^{19}$ distinct combinations instead of the original $2^{36}$.

Evaluations of the optimization problem \eqref{eq:trusstopo_socpmod2} for all the combinations took $9.5$ core hours. During the enumeration, a~globally optimal design of the complementary strain energy $c^* = 62.7$ was obtained, see Fig. \ref{fig:beam_global}, yielding a~$1.3\%$ increase of the objective function compared to the lower-bound design. The module set of the globally optimal design, shown in Fig. \ref{fig:beam_globaltileset}, contains $13$ modules that contribute to the load transfer, allowing for potential elimination of three empty modules from the set. Overall, the enumerated combinations generate a~nearly Gaussian distribution with a~mean value of $107.7$ and a~standard deviation of $14.6$, see Fig. \ref{fig:histogram}.

\begin{figure*}[!b]
	\begin{subfigure}{84mm}
		\includegraphics[width=\linewidth]{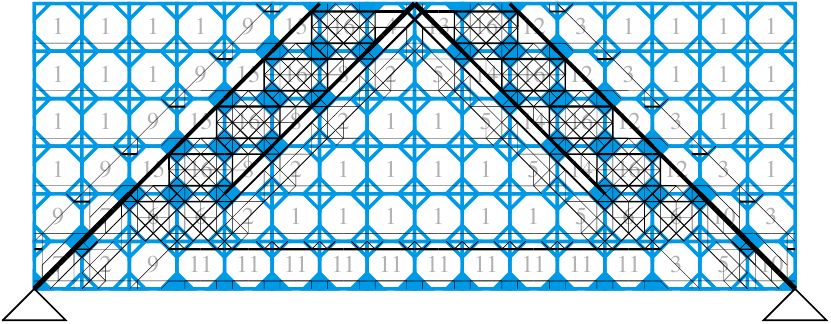}
		\caption{}
	\end{subfigure}\hfill%
	\begin{subfigure}{84mm}
		\vspace{5.4mm}
		\includegraphics[width=\linewidth]{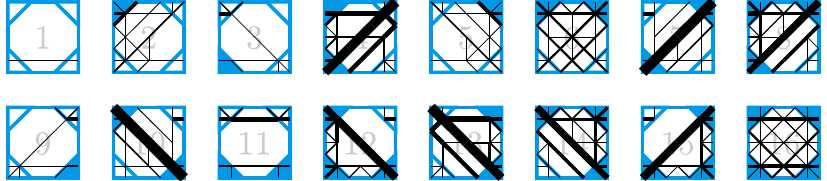}
		\vspace{5.4mm}
		\caption{}
	\end{subfigure}
	\caption{Best design of the (a) finely discretized beam with $c = 71.6$ as obtained via bilevel optimization, and (b) the corresponding tile set.}
	\label{fig:GAfiner_best}
\end{figure*}

\paragraph{Bilevel optimization using GA}

The bilevel optimization sol\-ver was launched $50$ times, each time with a~random population of $16$ individuals, evaluating the statistical properties of the bilevel optimization approach. The distribution of the complementary strain energy of the best individual within the population is shown in Fig. \ref{fig:GA}.

The initial random populations yielded topologies with a~mean energy of $107.4$, approximately matching the mean value $107.7$ of the nearly Gaussian distribution of the complete enumeration. Throughout the prescribed $40$ generations of the genetic algorithm, the complementary strain energy decreased to the final mean value of the best individual, $67.4$, being on average $8.9\%$ higher than the lower-bound solution and $7.5\%$ higher than the global optimum. Through bilevel optimization, a~second-best design, with strain energy $64.6$, was obtained. All the achieved objectives are within the lowest $0.2\%$ of all combinations, recall Fig.~\ref{fig:histogram}.

\subsubsection{Fine discretization}\label{sec:finely}

\begin{figure*}[!t]
	\setcounter{figure}{10}
	\centering
	\includegraphics[width=\linewidth]{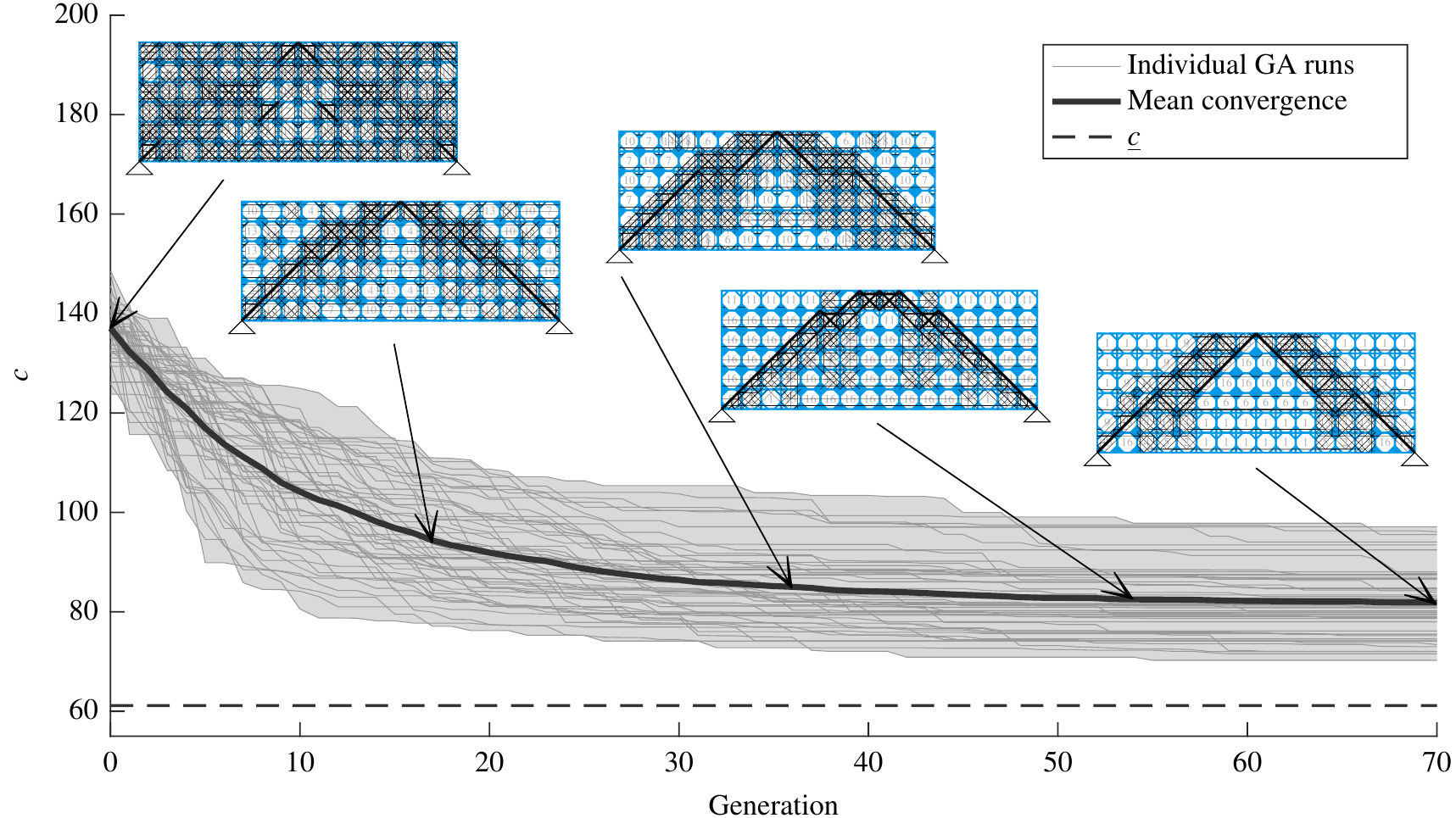}
	\caption{Convergence of complementary strain energies $c$ of the best individuals with $50$ independent runs of the genetic algorithm for the finely discretized beam; $\underline{c}$ stands for the complementary strain energy of the optimal non-modular design.}
	\label{fig:GAfiner}
\end{figure*}

Let us now consider a~beam with the same dimensions and with identical boundary conditions as in the previous subsection, recall Fig. \ref{fig:beam_dimensions}, but with a~refined discretization with $96$ modules, each with side lengths of $0.5$. To preserve comparability with the previous case, the connectivity matrix again satisfies symmetry along the midspan of the beam. Consequently, fine discretization permits $2^{62}$ distinct combinations of assemblies.

This huge number of combinations, pronounced further with an increased number of degrees of freedom, makes it impossible to perform the complete enumeration as in the previous case, leaving us without the knowledge of a~guaranteed global optimum. However, similarly to the previous example, we can obtain the bounds on the optimum: $\underline{c} = 61.1$ and $\overline{c} = 228.7$, implying that $c^* \in \left[61.1; 228.7 \right]$. 

\begin{figure}[!b]
	\setcounter{figure}{11}
	\centering
	\includegraphics[width=\linewidth]{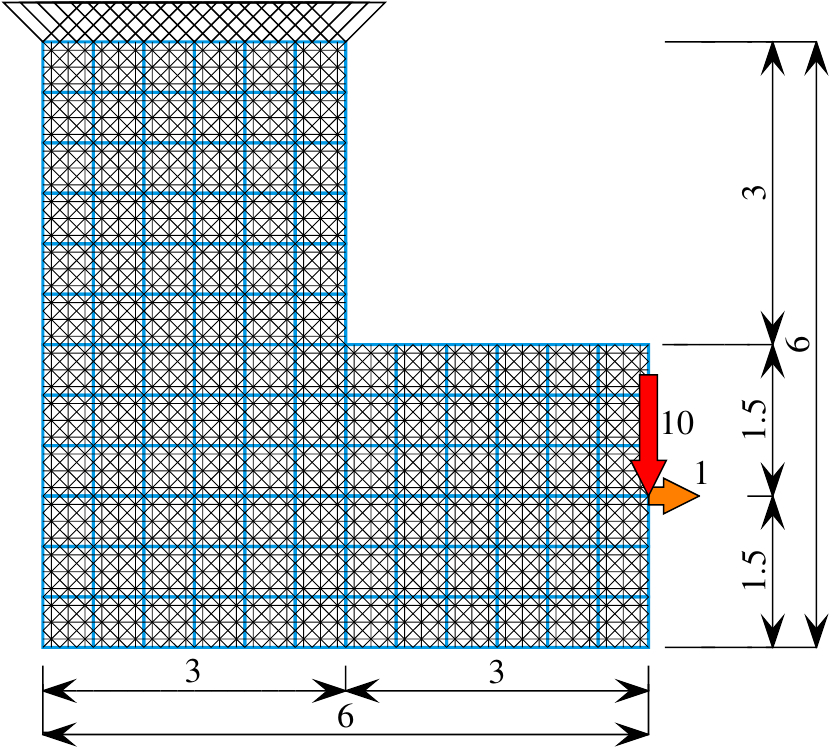}
	\caption{Dimensions, discretization into modules, boundary conditions, and ground structure of the L-shaped domain.}
	\label{fig:L_domain}
\end{figure}

Compared to the coarse problem, fine discretization produces a~richer ground structure, which allows the algorithm to reach a~decreased lower-bound complementary strain energy. Conversely, the upper-bound energy noticeably increases because a~larger ratio of material volume $\overline{V}$ appears to be placed inefficiently. Similar consequences of modularity also emerged in~\citep{Alexandersen:2015:TOM,Huang:2008:ODP}.

\begin{figure*}[!t]
	\setcounter{figure}{12}
	\centering
	\includegraphics[width=\linewidth]{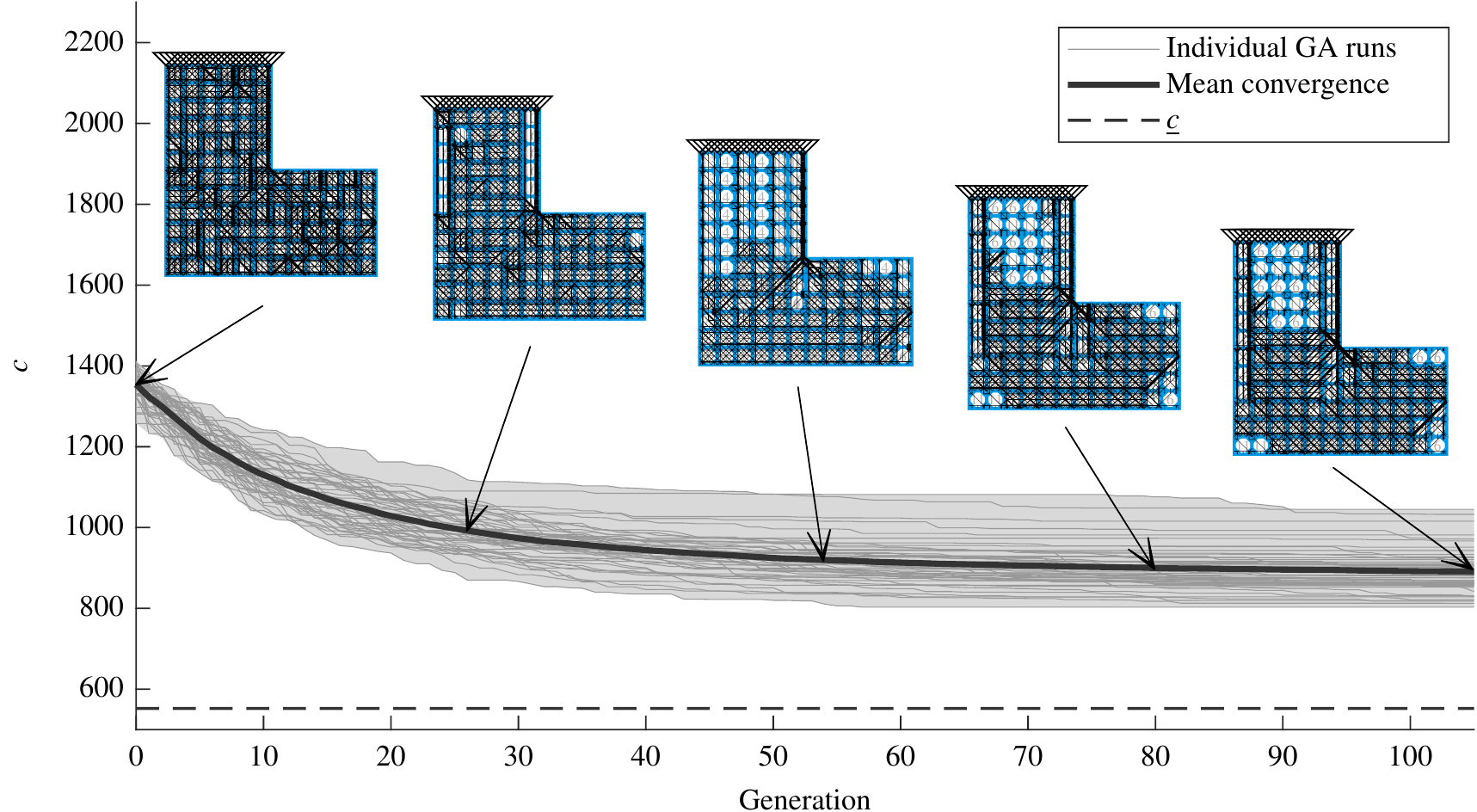}
	\caption{Convergence of complementary strain energies $c$ of the best individuals with $50$ independent runs of the genetic algorithm for an L-shaped beam; $\underline{c}$ stands for the complementary strain energy of the optimal non-modular design.}
	\label{fig:GA_L}
\end{figure*}

Convergence of the bilevel optimization algorithm, independently launched 50 times, is shown in Fig. \ref{fig:GAfiner}. The initial random populations of $29$ individuals determined designs of the mean objective value $156.5$---a~significant increase ($45.3\%$) compared to coarse discretization. Throughout $70$ generations, bilevel optimization converged to mean objective value of $82.1$, being $34.4\%$ more-compliant than the lower-bound design. The best design achieved, with $c=71.6$, is shown in Fig.~\ref{fig:GAfiner_best}, which amounts to a~$17.0\%$ increase over the lower-bound complementary strain energy $\underline{c}$.

\subsection{L-shaped beam with stress constraints and multiple load cases}\label{sec:L}

As the second illustrative problem, we assume an L-shaped design domain as shown in Fig.~\ref{fig:L_domain}. For this domain, two equally-weighted load cases, indicated by the two arrows in Fig.~\ref{fig:L_domain}, apply. Furthermore, we limit the structural volume by $\overline{V} = 100$, set the Young modulus to $E = 1$, and fix the maximum value of stress to $\sigma_\mathrm{UB} = -\sigma_\mathrm{LB} = 20$. Although the maximum (absolute value of) stress equals to $4.6$ in the lower-bound non-modular setting, which makes the stress constraint inactive, these constraints become active for some modular designs. For example, the worst-case modular design would yield a~maximum stress of $39.4$ without stress constraints. When imposed, the worst-case modular complementary strain energy approaches $\overline{c} = 1837.9$. Because $\underline{c} = 552.3$ arises from the lower-bound non-modular design, the optimum lies in $c^* \in [552.3, 1837.9]$.

After launching the bilevel optimization approach, the first generation of random individuals yielded a~mean energy of $1468.4$, a~value not too distant from the worst case. Through $106$ generations of $42$ individuals, the population evolved to set the mean energy of the best individuals at $891.0$, including a~best design of $803.1$ ($45.4\%$ higher than $\underline{c}$), and the worst design had a~complementary strain energy of $1045.0$. See Fig.~\ref{fig:GA_L} for the statistics of the best individuals within $50$ independent random runs of the algorithm.

\begin{figure*}[!b]
	\centering
	\setcounter{figure}{14}
	\begin{subfigure}{84mm}
		\includegraphics[width=\linewidth]{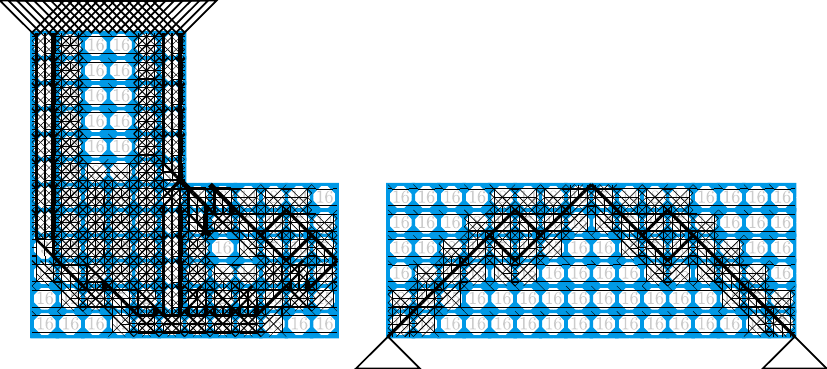}
		\caption{}
	\end{subfigure}\hfill%
	\begin{subfigure}{84mm}
		\vspace{7.8mm}
		\includegraphics[width=\linewidth]{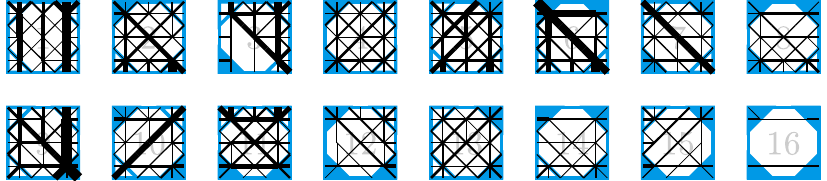}
		\vspace{7.8mm}
		\caption{}
	\end{subfigure}
	\caption{(a) Best reusable designs of the two domains with $c=829.6$ as obtained via bilevel optimization, and (b) the corresponding tile set.}
	\label{fig:GA_LandB_best}
\end{figure*}

\begin{figure}[!t]
	\setcounter{figure}{13}
	\begin{subfigure}{\linewidth}
		\includegraphics[width=\linewidth]{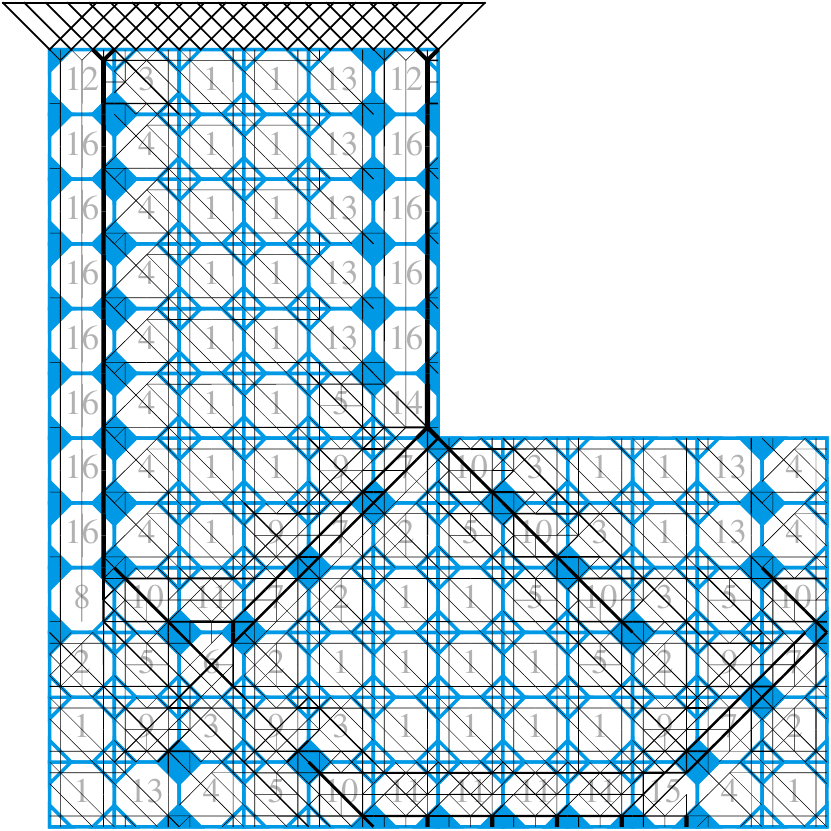}
		\caption{}
	\end{subfigure}\\
	\begin{subfigure}{\linewidth}
		\includegraphics[width=\linewidth]{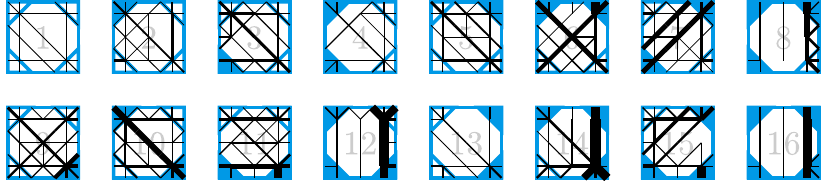}
		\caption{}
	\end{subfigure}
	\caption{Best design of the (a) evaluated L-shaped beam with $c=803.1$ as obtained via bilevel optimization, and (b) the corresponding tile set.}
	\label{fig:GA_L_best}
\end{figure}

While the best design, Fig.~\ref{fig:GA_L_best}, clearly aligns structural stiffness with the principal stress direction of the first load case (e.g., modules $10$, $11$, $14$, $16$), the interior of the module types $1$, $4$, and $13$ serves mainly as a~structural stabilization against the second load case.

\subsection{Module reusability in simply-supported and L-shaped beams}\label{sec:reusability}

The final example concerns the concurrent design of the finely discretized beam from Section \ref{sec:finely} with the L-shaped domain from Section \ref{sec:L}. In this case, the modules become reusable among these two domains, which is a~key benefit of modularity. Additionally, we introduce three minor deviations from the settings of the original problems: the stress constraints of the L-shaped domain also apply to the hinge-supported beam; the hinge supported beam does not enforce symmetric colorings; and instead of independent volume constraints, we constrain the overall volume by $\overline{V} = 200$. All these changes are justified by practical considerations: stress constraints should apply over the same material, symmetric coloring of the simply-supported beam may become inefficient as the L-shaped domain lacks the symmetry, and (one of the) volume constraints may become inactive. For these reasons, the designs in this section are not directly comparable with the previous ones. However, the original constraints can still be imposed, with few adjustments in the code.

In the modular-topology optimization framework, we obtained the lower-bound complementary strain energy of $\underline{c} = 490.5$ and a worst-case design with $\overline{c} = 2056.8$, implying that $c^* \in [490.5, 2056.8]$. In $50$ independent runs of the algorithm, the originally random population of $57$ individuals evolved in $140$ generations from an initial mean complementary strain energy of $1695.1$ to the best individual having the complementary strain energy of $924.5$, see Fig.~\ref{fig:GA_LandB}. The best design acquired, Fig.~\ref{fig:GA_LandB_best}, exhibited a~strain energy of $829.6$, which is $69\%$ worse than the lower-bound design. Modules in the design of Fig.~\ref{fig:GA_LandB_best}b are clearly distinguished by their major effective stiffness directions: the vertical direction of module $1$; the almost horizontal direction of modules $11$ and $16$; and the rest inclined ($2$, $3$, $5$, $6$, $7$, $9$, $10$, $12$, $15$), or stiff in three ($8$, $14$) or all directions ($4$, $13$).

\begin{figure*}[!t]
	\centering
	\setcounter{figure}{15}
	\includegraphics[width=\linewidth]{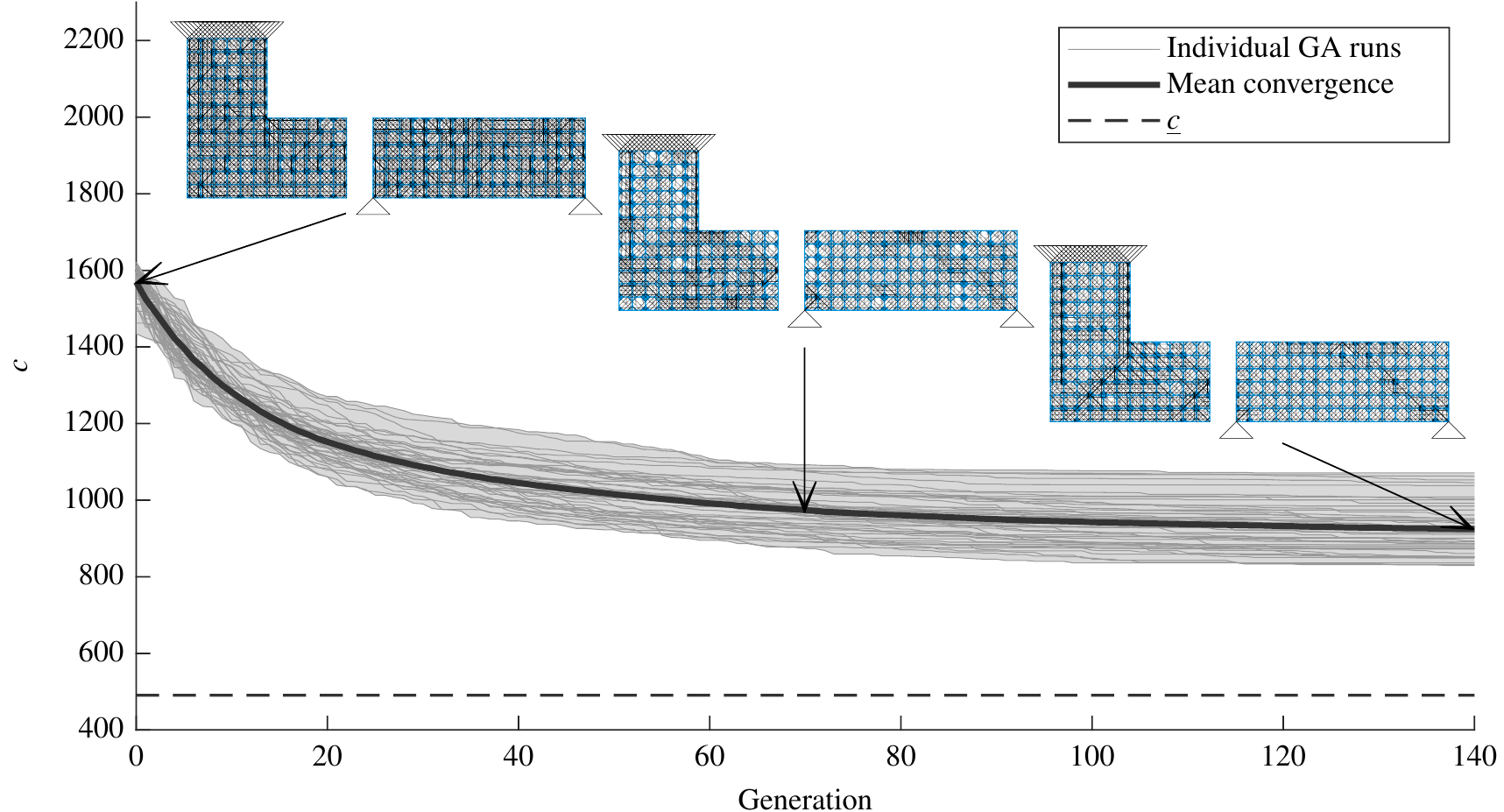}
	\caption{Convergence of complementary strain energies $c$ of the best individuals with $50$ independent runs of the genetic algorithm; $\underline{c}$ stands for the complementary strain energy of the optimal non-modular design.}
	\label{fig:GA_LandB}
\end{figure*}

\section{Conclusions}

In this paper, we introduced a~novel bilevel modular-topology optimization approach, facilitating simultaneous optimization of the topologies of $16$ independent truss modules together with their optimal placement within their respective structural macro-scale domains. This method adopts the concept of corner Wang tiles as a~suitable formalism for describing non-periodic assemblies of structural modules to maintain edge compatibility: a new class of compatible and reconfigurable \mbox{(micro-)}structures.

Lower-level optimization constitutes the truss topology least-compliant design problem, extended to structural modularity, stress constraints, multiple load cases, and modules reusability. We formulate this optimization problem as a~convex second-order cone program (SOCP), efficiently solvable to global optimality by employing modern mathematical programming solvers. In addition, modularity enables us to aggregate the constraints and variables of the original non-modular problem, resulting in the optimal modular designs being found faster compared to their non-modular counterparts. Since the compliances of modular designs are bounded by the periodic-unit-cell (PUC) design from above and by the non-modular design from below, any (and even random) assembly plan balances the solution efficiency of PUC with the design performance of the non-modular design. The final optimized design quality thus strongly depends upon the supplied assembly plan. To mitigate and take advantage of this dependence, we have developed a~bilevel modular-topology optimization framework, using meta-heuristics (namely, a~genetic algorithm) to search for an optimal assembly plan.

After implementing this approach in \textsc{Matlab}, we assess its performance using three sample problems. For the first, we consider a~hinge-supported beam. For the case illustrating coarse discretization and symmetric module interfaces, we compute the globally optimal design using brute-force enumeration. It turns out that the optimal modular design almost achieves the quality of a~non-modular one. The bilevel optimization approach converges to solutions near the optimum, and is, therefore, suitable for finding an approximate solution to the optimization problem in much shorter times compared to the enumeration. When a~finer discretization for the hinge-supported beam is adopted, the quality of the optimized modular design decreases compared to coarse discretization. This issue seems to be a~common drawback of modularity, as reported earlier by \citet{Huang:2008:ODP}. For the second and third problems, which include an L-shape design domain, we demonstrated how stress constraints, multiple load cases, and module reusability can be imposed while maintaining convexity (and thus solution efficiency) for the inner optimization problem. For both of these sample problems, the optimized designs outperform the optimal PUC designs considerably, and exhibit structural efficiency and material distribution nearly equivalent to the non-modular designs.

In the future, several important extensions need to be considered to build upon these pilot results. First, upper-level optimization could be replaced by a~heuristic procedure based on free-material optimization \citep{Zowe1997} or by machine learning. Machine learning may also allow for us to design the topology of individual modules \citep{Gu2018}. Second, since the chosen Wang tileset plays a~crucial role, different tilesets may yield different optimal designs. Adaptive choice of a~set cardinality and distribution of color codes within set is worth investigating. Similarly, the module ground structure as well as the shape of the modules influence the final design; for instance, in the first example of the hinge-supported beam, the best modular design always tends to be the two-bar truss when a~correct aspect ratio of the modules and a~suitable module ground structure is used. 

The proposed strategy readily extends to three dimensions using Wang cubes in $3$D \citep{Culik1996}, so that optimized module sets could be applied in the modular design of $3$D-printed \textsc{Lego}\textsuperscript{\textregistered}-like products~\citep{Schumacher:2015:MCE} or combinatorial aperiodic metamaterials~\citep{Coulais:2016:CDM} with complex shapes~\citep{Antolin2019}. Moreover, the inner SOCP formulation also allows for other convex extensions, e.g., the fundamental free-vibrations eigenfrequency lower-bound constraint~\citep{Ohsaki1999,Tyburec:2019:DM3} and bounds on peak power~\citep{Heidari2009}. An extension to modular buckling mechanisms \citep{Oliveri2020} is another challenge. Last, adopting continuum topology optimization might also provide invaluable insights and broaden the potential of the proposed approach.

\begin{acknowledgements}
We thank Michal Ko\v{c}vara for valuable suggestions, and Stephanie Krueger for a critical review of the initial versions of this manuscript.
\end{acknowledgements}

{\small\noindent \textbf{Funding}\hspace{1mm} The authors acknowledge the financial support of the Czech Science Foundation project No. 19-26143X.}

\section*{Compliance with ethical standards}

{\small\noindent \textbf{Conflict of interest}\hspace{1mm} The authors declare that they have no conflict of interest.}

{\small\noindent \textbf{Replication of results}\hspace{1mm} Source codes are available at \citep{tyburec_marek_2020_3750751}.}

\appendix

\section{Complementary strain energy conic constraints for modular designs}\label{app:SOCP}

Although the original elastic design formulation \eqref{eq:trusstopo_nonlin} lacks convexity, it allows for reformulation into a~convex conic optimization problem. Here, we consider dual complementary-strain-energy reformulation \eqref{eq:trusstopo_socp} that can exploit modularity by aggregation of constraints and design variables. This section of the appendix derives this aggregation and explains the basic mechanical reasoning behind the reformulation.

Let us assume a~minimization of the structural complementary strain energy function, Eq. \eqref{eq:trusstopo_socp_compl}, defined to be the sum of the upper bounds for the complementary strain energies $w_i$ of individual bars $i \in \{1\;..\;n_\mathrm{b}\}$, i.e.,
\begin{equation}
w_i \ge \frac{1}{2} \frac{\ell_i}{E_i} \frac{s_i^2}{a_i},
\end{equation}
where $\ell_i$, $E_i$, and $a_i$ are the length, Young modulus, and the cross-section area of the $i$-th element, respectively. The axial force in this element is denoted by $s_i$. Notice that since we minimize the sum of the upper bounds $w_i$ \eqref{eq:trusstopo_socp_compl}, they attain the value of the complementary strain energy at the optimum, which is in turn equal to the compliance in \eqref{eq:trusstopo_nonlin}.

Instead of minimizing the sum of complementary strain energies of individual bars we can, however, minimize the sum of aggregated complementary strain energies
\begin{equation}
w_{\mathrm{g},j} = [\textbf{G}_{:,j} (\mathbf{C})]^\mathrm{T} \mathbf{w} \ge \frac{1}{2 a_{\mathrm{g},j}} [\textbf{G}_{:,j} (\mathbf{C})]^\mathrm{T} \left(\bm{\ell} \oslash \mathbf{E} \odot \mathbf{s}^{\circ 2}\right),\label{eq:cc_sum}
\end{equation}
where $w_{\mathrm{g},j}$ is the upper bound for the sum of complementary strain energies of the bars that share the cross-section area $a_{\mathrm{g},j}$, and $\oslash$, $\odot$ with $^{\circ}$ are the Hadamard ``element-wise'' division, multiplication, and power. This step effectively eliminates summands in the objective function and aggregates constraints as well as design variables.

Because \eqref{eq:cc_sum} is not defined for $a_{\mathrm{g},j}=0$, we perform a~multiplication by the non-negative $4 a_{\mathrm{g},j}$ to obtain
\begin{equation}\label{eq:zero}
4 w_{\mathrm{g},j} a_{\mathrm{g},j} \ge [\textbf{G}_{:,j}(\mathbf{C})]^\mathrm{T} \left( 
\left[2 \bm{\ell} \oslash \mathbf{E}
\right]^{\circ\frac{1}{2}} \odot \mathbf{s}\right)^{\circ2}.
\end{equation}
Eq.~\eqref{eq:zero} now allows for zero cross-section areas as is required by topology optimization. However, in this case, the corresponding internal forces vanish and $w_{\mathrm{g},j}$ is arbitrary. Notice that $w_{\mathrm{g},j}$ may even attain arbitrarily low negative values, making the complementary strain energy functional non-physical and the objective function \eqref{eq:trusstopo_socp_compl} unbounded.

Because the aggregated constraints share the same cross-section, adding $w_{\mathrm{g},j}^2 - 2 w_{\mathrm{g},j} a_{\mathrm{g},j} + a_{\mathrm{g},j}^2$ to both sides of the inequality provides us with the sum-of-squares inequality
\begin{equation}
\left(w_{\mathrm{g},j} + a_{\mathrm{g},j} \right)^2 \ge \left(w_{\mathrm{g},j} - a_{\mathrm{g},j}\right)^2 
+
[\textbf{G}_{:,j} (\mathbf{C})]^\mathrm{T} \left( 
\left[2 \bm{\ell} \oslash \mathbf{E}
\right]^{\circ\frac{1}{2}} \odot \mathbf{s}\right)^{\circ2},\label{eq:cc_2}
\end{equation}
which is equivalent to 
\begin{subequations}
\begin{align}
w_{\mathrm{g},j}^+ + a_{\mathrm{g},j}^+ \ge \left\lVert 
\begin{pmatrix}
w_{\mathrm{g},j}^+ - a_{\mathrm{g},j}^+\\
\textbf{G}_{:,j}(\mathbf{C}) \odot \left[2 \bm{\ell} \oslash \mathbf{E}
\right]^{\circ\frac{1}{2}} \odot \mathbf{s}
\end{pmatrix}
\right\rVert_2,\\
-w_{\mathrm{g},j}^- - a_{\mathrm{g},j}^- \ge \left\lVert 
\begin{pmatrix}
w_{\mathrm{g},j}^- - a_{\mathrm{g},j}^-\\
\textbf{G}_{:,j}(\mathbf{C}) \odot \left[2 \bm{\ell} \oslash \mathbf{E}
\right]^{\circ\frac{1}{2}} \odot \mathbf{s}
\end{pmatrix}
\right\rVert_2,\label{eq:socp_app2}
\end{align}
\end{subequations}
with 
\begin{subequations}
\begin{align}
a_{\mathrm{g},j} &=  a_{\mathrm{g},j}^+ - a_{\mathrm{g},j}^-,\\
w_{\mathrm{g},j} &=  w_{\mathrm{g},j}^+ - w_{\mathrm{g},j}^-,\\
a_{\mathrm{g},j}^+&\ge 0, a_{\mathrm{g},j}^-\ge 0, w_{\mathrm{g},j}^+ \ge 0 ,w_{\mathrm{g},j}^-\ge 0.
\end{align}
\end{subequations}
Clearly, \eqref{eq:socp_app2} is redundant, as both the $a_{\mathrm{g},j}$ and $w_{\mathrm{g},j}$ must be non-negative, i.e., $w_{\mathrm{g},j}^- = a_{\mathrm{g},j}^- = 0$. Moreover, if $\mathbf{a}_\mathrm{g} \ge \mathbf{0}$ is enforced explicitly, which is our case---recall Eq. \eqref{eq:trusstopo_socp_areas}---the non-physical situation of the negative complementary strain energy is automatically eliminated because the Euclidean norm is non-negative by definition. Consequently, we end up with the conic constraint
\begin{equation}
w_{\mathrm{g},j} + a_{\mathrm{g},j} \ge \left\lVert 
\begin{pmatrix}
w_{\mathrm{g},j} - a_{\mathrm{g},j}\\
\textbf{G}_{:,j}(\mathbf{C}) \odot \left[2 \bm{\ell} \oslash \mathbf{E}
\right]^{\circ\frac{1}{2}} \odot \mathbf{s}
\end{pmatrix}
\right\rVert_2,\label{eq:cc_fin}
\end{equation}
which is convex and equivalent to \eqref{eq:cc_sum} for all positive cross sections. For zero cross sections, complementary strain energy is implicitly enforced to be non-negative, and actually zero, since $w_{\mathrm{g},j}$ is to be minimized.

\section{Binary genetic algorithm}\label{app:GA}

In the considered bilevel optimization problem, we use the following parameters for the genetic algorithm: The population consists of $n_\mathrm{pop}$ individuals, heuristically set to
\begin{equation}
n_\mathrm{pop} = \left\lfloor 3.6 \sqrt{\lvert \mathbf{C} \rvert}+ 0.5
\right\rfloor,
\end{equation}
where $\lfloor \bullet \rfloor$ denotes rounding of $\bullet$ towards the nearest integer less than or equal to $\bullet$.

The individuals evolve through $n_\mathrm{gen}$ generations, where
\begin{equation}
n_\mathrm{gen} = 5 \left\lfloor 0.49 n_\mathrm{pop}+ 0.5
\right\rfloor.
\end{equation}
Further, the \textit{selection} of parents giving birth to offspring follows from tournament selection. The size of tournament $n_\mathrm{t}$ equals
\begin{equation}
n_\mathrm{t} = \left\lfloor\frac{4}{3} \sqrt{\lvert \mathbf{C} \rvert}+0.5\right\rfloor,
\end{equation}
and the individuals participating in the tournament are chosen randomly. Sorted accordingly to their fitness values, the probability $p_i$ of the $i$-th individual to win the tournament equals
\begin{equation}
p_i = p_\mathrm{t} \cdot (0.7)^{i},
\end{equation}
where $p_\mathrm{t} = 0.3$.

For combinations of individuals, we used uniform \textit{cross-over} with a~probability of $p_\mathrm{c} = 0.94$ and a~combination of the parents genes based on their fitness, or keeping the better parent otherwise.

The \textit{mutation} operator applies for each gene with a~probability of 
\begin{equation}
p_\mathrm{m} = \frac{1}{\lvert \mathbf{C} \rvert},
\end{equation}
reversing the binary value of the affected genes. Additionally, the genetic algorithm was set to guarantee population \textit{diversity}, i.e., substituted duplicate individuals with random ones and to keep the best individual through \textit{elitism}.

\bibliography{liter_abbr}
\bibliographystyle{springernat}

\end{document}